\documentclass[12pt]{article}
\def\half{{\textstyle{1\over2}}}
\def\2{{1\over 2}}

\newcommand{\rf}[1]{(\ref{#1})}
\def\b{\bar}

\renewcommand{\t}{\tilde}
\newcommand{\p}{\partial}
\newcommand{\bp}{\bar{\partial}}
\newcommand{\tc}{\tilde c }
\usepackage{amssymb}

\title{
\bf{Formal Maurer-Cartan Structures: from CFT to Classical Field Equations}}
\author{Anton M. Zeitlin\footnote{anton.zeitlin@yale.edu,\newline
 http://pantheon.yale.edu/$\sim$az84, http://www.ipme.ru/zam.html}   
\footnote{On leave of absence from the St.-Petersburg Division of Steklov Mathematical Institute}\\
Department of Mathematics,\\
Yale University,\\
442 Dunham Lab, 10 Hillhouse Ave\\
New Haven, CT 06511}
\date{}

\begin{document}
\maketitle
\begin{abstract}
We show how the well-known classical field equations 
as Einstein and Yang-Mills ones, which arise as the conformal invariance conditions of certain two-dimensional theories, 
expanded up to the second order in the formal parameter, can be reformulated as Generalized/formal Maurer-Cartan equations 
(GMC), where the differential is the BRST operator of String theory. We introduce the bilinear operations which 
are present in GMC, and study  their properties, 
allowing us to find the symmetries of the resulting equations which will be naturally identified 
with the diffeomorphism and gauge symmetries of Einstein and Yang-Mills equations correspondingly. 
\end{abstract}
\section{Introduction}
It is well known from the context of String Field Theory (SFT) \cite{siegel} 
that the {\it linearized} versions of classical field equations such as Einstein and Yang-Mills ones appear to be 
the closeness conditions of certain operators (depending on both the ``matter'' fields and $c$-ghosts) 
with respect to the BRST operator \cite{brst}, \cite{pol}:
\begin{eqnarray}\label{exact}
[Q,\phi^{(0)}]=0.
\end{eqnarray} 
One might expect, following further the canonical construction of  closed SFT \cite{zwiebach} that the generalization of 
\rf{exact} which corresponds to {\it nonlinear} field equations, should be of the form of Generalized/formal Maurer-Cartan equation 
\cite{losev}:
\begin{eqnarray}\label{genmc}
[Q,\phi^{(0)}]+C_2(\phi^{(0)},\phi^{(0)})+ C_3(\phi^{(0)},\phi^{(0)},\phi^{(0)})+...=0,
\end{eqnarray} 
where $C_n$ are graded (w.r.t. the ghost number) multilinear operations, satisfying the certain quadratic relations leading to 
the homotopic Lie algebra \cite{zwiebach}, \cite{stasheff}. In this paper, we formulate it only as a {\it hypothesis}, namely, 
we consider only the second order corrections to this equation, therefore, we give an explicit construction only of 
operation $C_2$ 
and verify the relations between $Q$ and $C_2$. We postpone the proof of the quadratic relations between higher $C_n$- operations 
(homotopy Jacobi identity and other relations of $L_{\infty}$ algebra) 
until we will be interested in the higher order corrections to equation \rf{genmc} \footnote{We indicate here that in the case of 
Yang-Mills equations this proof is already given on the field theory level \cite{zeit3}: the operations $C_2$ and $C_3$ are shown 
to satisfy the homotopy Jacobi identity and other relations necessary for the homotopy Lie algebra. All other operations 
are equal to zero in this case.  }. We will consider them in the further publications on 
the subject.  

The nonlinear field equations under discussion (Einstein and Yang-Mills ones) 
correspond to the 1-loop conformal invariance conditions in the certain perturbed two-dimensional conformal field theories. 
Therefore, we expect that the explicit operator formulation of  operations $C_n$ will include further corrections and 
lead to the precise meaning of the beta-function in conformal perturbation theory (this problem was already mentioned
 in the context of SFT 
\cite{sen}). 

In \cite{zeit2}, we considered the perturbed $\beta$-$\gamma$ system and constructed the $C_2$ operation. In that case, the one-loop 
beta function was bilinear in the perturbation operator, therefore, the second order approximation of \rf{genmc} 
gave exact results. In this paper, we continue the consideration of the second order approximation of \rf{genmc}, namely:
\begin{eqnarray}\label{genmcsec}
[Q,\phi_1^{(0)}]=0, \quad [Q,\phi_2^{(0)}]+C_2(\phi_1^{(0)},\phi_1^{(0)})=0.
\end{eqnarray} 
Here, we have expanded $\phi^{(0)}=\sum^{\infty}_{n=1}t^n\phi_n^{(0)}$ with respect to the formal parameter $t$ and in such a way, 
\rf{genmcsec} corresponds to the first and the second order of the expansion of \rf{genmc}. In section 2,
 we consider operation $C_2$ and introduce some of its
properties, which will allow us to formulate the operator symmetries of \rf{genmc} and 
to relate \rf{genmcsec} with the second order approximation of the 
equations of the conservation of deformed BRST current in perturbed theory, considered in \cite{zeit}. In  subsection 2.3,
we  study an example which is related to the string theory in background of metric and dilaton described by the 
sigma model (see e.g. \cite{tseytlin}):
\begin{eqnarray} \label{sma}
S=\int_{\Sigma} d^2 z (\frac{1}{4\pi\alpha'}G_{\mu\nu}(X)\p X^{\mu}\bp X^{\nu}+\frac{1}{2}\sqrt{\gamma}R^{(2)}(\gamma)\Phi(X)).
\end{eqnarray}
It is well known \cite{eeq}-\cite{Polbook} that the equations of conformal invariance for the model \rf{sma} are Einstein equations:
\begin{eqnarray}\label{gdil}
&&R_{\mu\nu}+2\nabla_{\mu}\nabla_{\nu}\Phi=0,\nonumber\\
&&R+4\nabla_{\mu}\nabla^{\mu}\Phi-4\nabla^{\mu}\Phi\nabla_{\mu}\Phi=0.
\end{eqnarray}
We show that  equations \rf{genmcsec} reproduce, at the lowest orders in $\alpha'$, 
 equations \rf{gdil} up to the second order of expansion of the fields: 
$G_{\mu\nu}=\eta_{\mu\nu}-th_{\mu\nu}(X)-t^2s_{\mu\nu}(X)+O(t^3)$, $\Phi=t\Phi_1+t^2\Phi_2+O(t^3)$. We also demonstrate that the 
operator symmetries of \rf{genmcsec} correspond to the diffeomorphism symmetries of equations \rf{gdil}. 
However, there is a certain ambiguity, 
since the choice of constant metric $\eta_{\mu\nu}$ and the deformation 
parameter $t$ is definitely not unique. This is a common problem (and it is not our aim in this paper to get rid of it), 
when one wants to consider the perturbation theory for the sigma models, 
where it is impossible to extract the free action without destroying the 
geometric context. In such a way, the first order formulation of the sigma-model, 
introduced in \cite{zeit2},\cite{lmz}, looks more promising from 
this point of view. 

In section 3, we consider the boundary CFT corresponding to the open string on the disc, conformally mapped to the half-plane. 
We introduce the bilinear operation $C_2$ which now is a bilinear operation on the space of 
tensor product of CFT operators with some 
Lie algebra, and consider the analogy of equations \rf{genmcsec}. It appears that one can deduce the Yang-Mills equations: 
\begin{eqnarray}
\p_{\mu}F^{\mu\nu}+[A_{\mu},F^{\mu\nu}]=0,\quad F_{\mu\nu}=\p_{\mu}A_{\nu}-\p_{\nu}A_{\mu}+[A_{\mu},A_{\nu}]
\end{eqnarray}
up to 
the second order in the formal parameter (namely, if one considers the gauge field of the form  
$A_{\mu}=tA^1_{\mu}+t^2A^2_{\mu}+O(t^3)$). 
It is interesting to note that the usual construction of the open SFT \cite{thorn} is related to homotopy associative 
algebra $A_{\infty}$ \cite{stasheff}, generated by Witten's product \cite{witten}. If our conjectures are correct, 
in the case of open string there also exists  the structure of homotopy Lie algebra. 
 
In Conclusion, we give final remarks and mention the ways of further development of this formalism.

\section{CFT, Closed Strings in Background Fields and Einstein Equations}
\subsection*{Notation and conventions} 
Throughout this section, we assume that all matter field operators have the operator products of the following form:
\begin{eqnarray}\label{ope}
&&V(z)W(z')=\sum_{r=-\infty}^{m}\sum_{s=-\infty}^{n}(V,W)_p^{(r,s)}(z')\nonumber\\
&&(z-z')^{-r}(\b z-\b z')^{-s}(\log(|(z_1-z_2)/\mu|)^p,
\end{eqnarray} 
where $\mu$ is some parameter. We consider the ghost fields 
$b(z)$, $c(z)$ and $\t b(\b z)$, $\t c(\b z)$ of conformal weights $(2,0), (-1,0)$ and  $(0,2), (0,-1)$ 
correspondingly, which have the following operator products:
\begin{eqnarray}
b(z)c(w)\sim \frac{1}{z-w},\quad \t b(\b z) \t c(\b w)\sim \frac{1}{\b z-\b w}.
\end{eqnarray} 
The so-called ghost number operator is of the following form: 
\begin{eqnarray}\label{gn}
N_g=\int (dzj_g-d\b z \t j_g),  
\end{eqnarray}
where $j_g=-bc$ and $\t j_g=-\t b \t c$. For the given conformal field theory with the holomorphic and antiholomorphic components of 
energy-momentum tensor $T(z)$, $\t T(\b z)$, one can define a BRST operator:
\begin{eqnarray}
\label{BRST}
&&Q=\frac{1}{2\pi i}\oint\mathcal{J_B},\quad \mathcal{J_B}=j_Bdz-\tilde{j_B}d\bar{z},\\
&&j_B=cT+:bc\partial c:, \quad \tilde{j}_B=\tilde{c}\tilde{T}
+:\tilde{b}\tilde{c}
\bar{\partial} \tilde{c}:.\nonumber
\end{eqnarray}
It is well known that this operator becomes nilpotent when the central charges in both holomorphic and 
antiholomorphic sectors of the theory are equal to 26.
 
We couple $c, \tc$ ghost fields to matter fields and denote the resulting space, that is the space of differential 
polynomials in $c, \tc$-ghosts with matter fields as coefficients, as $H^0$. 
If $\phi^{(0)}\in H^0$ is the eigenvector of the operator $N_g$ with the eigenvalue $n_{\phi}$, we 
say that this field is of ghost number $n_{\phi}$ (it is obvious that it can be only nonnegative integer), in other words, the space $H^0$ is graded with respect 
to $N_g$. 
It is also reasonable to define the spaces $H^1$, $H^2$ of 1-forms $\psi^{(1)}=\psi(z)dz-\b \psi(z)d\b z$ and 
2-forms $\chi^{(2)}=dz\wedge d\b zV_{\chi}$ (such that $\psi, \b \psi, V_{\chi}\in H^0$). Moreover, 
we associate with any field $\phi^{(0)}\in H^0$ the 
following 1-form and 2-form: 
\begin{eqnarray}
\phi^{(1)}=dz[b_{-1},\phi^{(0)}]+d\b z[\t b_{-1},\phi^{(0)}], \quad
\phi^{(2)}=dz\wedge d\b z[b_{-1},[\t b_{-1},\phi^{(0)}]],  
\end{eqnarray}
which satisfy the following $descent$ equations:
\begin{eqnarray}\label{desc}
[Q,\phi^{(1)}]=d\phi^{(0)}-[Q,\phi^{(0)}]^{(1)},\quad  [Q,\phi^{(2)}]=d\phi^{(1)}+[Q,\phi^{(0)}]^{(2)}.
\end{eqnarray}
In the following, we  use the notation $\p=\frac{\p}{\p z}$, $\bp=\frac{\p}{\p \b z}$. 

\subsection{Bilinear operation}
In this subsection, we discuss the bilinear operation which will be present in the generalized Maurer-Cartan 
equation, and relate it to the bilinear operation given in \cite{zeit}. 

First of all, we note the following. The object 
\begin{eqnarray}
\int_{C_{\epsilon,z}}dwV(w) W(z),
\end{eqnarray}
where $C_{\epsilon,z}$ is a circle contour of radius $\epsilon$ around point $z$ and $V,W$ are some operators, according to 
\rf{ope} belongs to the space of power series in $\epsilon$ and $\log\epsilon/\mu$. This gives us a possibility 
to write down the following definition.

\vspace*{3mm}

\noindent{\bf Definition 2.1.} {\it For any two operators $\phi^{(0)}$, $\psi^{(0)}\in H^0$ we define a bilinear operation
$M:H^0 \otimes H^0\to H^0$
\begin{eqnarray}
&&M(\phi^{(0)}, \psi^{(0)})(z)=\\
&&\frac{1}{4\pi i}\mathcal{P}\int_{C_{\epsilon,z}}\phi^{(1)}\psi^{(0)}(z)+(-1)^{n_{\phi}n_{\psi}}
\frac{1}{4\pi i}\mathcal{P}\int_{C_{\epsilon,z}}\psi^{(1)}\phi^{(0)}(z),\nonumber
\end{eqnarray}

\vspace*{3mm}

\noindent
where $\mathcal{P}$ is a projection on the $\epsilon^0(\log\epsilon/\mu)^0$ term. }

\vspace*{3mm}

\noindent
It is interesting to see how this operation behaves under the action of the BRST operator. The result is given by the following proposition.

\vspace*{3mm}

\noindent
 {\bf Proposition 2.1.} {\it Operation $M$ satisfies the relation:
\begin{eqnarray}\label{2prod}
[Q,M(\phi^{(0)}, \psi^{(0)})]+M( [Q,\phi^{(0)}], \psi^{(0)})+(-1)^{n_{\phi}}M( \phi^{(0)}, [Q,\psi^{(0)}])=0.
\end{eqnarray}}
{\bf Proof.} First, we need to show that BRST operator commutes with projection operator $\mathcal{P}$. Really, let's denote 
\begin{equation}
f(V,W)(z)=\int_{C_{\epsilon,z}}dwV(w)W(z)
\end{equation}
for some operators $V$, $W$. 

From \rf{ope} we know that 
$f(V,W)=\sum^{\infty}_{n=-k}\sum^{\infty}_{m=0}f_{n,m}(V,W)\epsilon^n(\log(\epsilon/\mu))^m$. 
The projection operator acts as follows:
$\mathcal{P}f(V,W)=f_{0,0}(V,W)$. Therefore we see that 
\begin{equation}
\mathcal{P}[Q,f(V,W)]=[Q,\mathcal{P}f(V,W)]=[Q,f_{0,0}(V,W)]. 
\end{equation}
In such a way we see that BRST operator commutes with projection operator and hence the relation \rf{2prod} 
can be easily established by means of the simple formula 
$[Q, \phi^{(1)}]=d\phi^{(0)}-[Q, \phi^{(0}]^{(1)}$. $\blacksquare$

\vspace*{3mm}

\noindent
{\bf Remark.} If one denotes $C_1=Q$ and $C_2=M$, 
from the Proposition 2.1. we get that the relations between operations $C_1$ and $C_2$ are as follows:
\begin{eqnarray}
&&C_1(C_1(\phi^{(0)}))=0, \\ 
&&C_1(C_2(\phi^{(0)}, \psi^{(0)}))+C_2( C_1(\phi^{(0)}), \psi^{(0)})+(-1)^{n_{\phi}}C_2( \phi^{(0)}, C_1(\psi^{(0)}))=0\nonumber
\end{eqnarray}
for any $\phi^{(0)},\psi^{(0)}\in H^0$. These formulas repeat the corresponding relations of homotopy Lie algebra 
\cite{zwiebach}, \cite{stasheff}.\\ 
Now we define another bilinear operation.

\vspace*{3mm}

\noindent
{\bf Definition 2.2.} {\it For any two fields $\phi^{(0)}$, $\psi^{(0)}\in H^0$ we define a bilinear operation
$K:H^0 \otimes H^0\to H^2$
\begin{eqnarray}
&&K(\phi^{(0)}, \psi^{(0)})(z)=\\
&&\frac{1}{2\pi i}\mathcal{P}\int_{C_{\epsilon,z}}\phi^{(1)}\psi^{(2)}(z)+(-1)^{n_{\phi}n_{\psi}}
\frac{1}{2\pi i}\mathcal{P}\int_{C_{\epsilon,z}}\psi^{(1)}\phi^{(2)}(z),\nonumber
\end{eqnarray}
where $\mathcal{P}$ is a projection on the $\epsilon^0(\log\epsilon/\mu)^0$ term. }

\vspace*{3mm}

\noindent
{\bf Remark.} This operation $K$ is the projected version of the operation $K_{\epsilon}$ defined in \cite{zeit}.\\
The properties of this operation are summarized in the following proposition.

\vspace*{3mm}

\noindent
 {\bf Proposition 2.2.}\\
1){\it Operation $M$ is related to operation $K$ in the following way:\\
\begin{eqnarray}
M(\phi^{(0)}, \psi^{(0)})^{(2)}=K(\phi^{(0)}, \psi^{(0)})+d\chi^{(1)},
\end{eqnarray}
where as usual $M(\phi^{(0)}, \psi^{(0)})^{(2)}=dz\wedge d\b z[b_{-1},[\t b_{-1},M(\phi^{(0)}, \psi^{(0)})]]$, 
 $\chi^{(1)}\in H^1$, and $d$ is the de Rham differential.}\\
2){\it Operation $K$ satisfies the following relation:
\begin{eqnarray}
[Q,K(\phi^{(0)}, \psi^{(0)})]+K( [Q,\phi^{(0)}], \psi^{(0)})+(-1)^{n_{\phi}}K( \phi^{(0)}, [Q,\psi^{(0)}])=d\lambda^{(1)}
\end{eqnarray}
for some $\lambda^{(1)}\in H^1$.}\\ 
To prove this proposition we need two lemmas.

\vspace*{3mm}

\noindent
{\bf Lemma 1.} {\it Consider $V, W\in H^0$. Then the expressions 
\begin{eqnarray}
\label{hol}f_1(V,W)(z)=\int_{C_{\epsilon,z}}dwV(w)W(z)+(-1)^{n_Vn_W}\int_{C_{\epsilon,z}}dwW(w)V(z),\\
\label{ahol}f_2(V,W)(z)=\int_{C_{\epsilon,z}}d\b wV(w)W(z)+(-1)^{n_Vn_W}
\int_{C_{\epsilon,z}}d\b wW(w)V(z)
\end{eqnarray} 
can be represented in the following form:
\begin{eqnarray}
f_i(V,W)(z)=\p \b g_i(V,W)(z)+\bp g_i(V,W)(z)
\end{eqnarray} 
for some operators $g_i,\b g_i\in H^0$, constructed from $(V,W)_k^{(r,s)}$ and their derivatives.}

\vspace*{3mm}

\noindent
The proof can be easily obtained using  \rf{ope} and comparing the coefficients 
$(V,W)_k^{(r+1,r)}$ and $(W,V)_k^{(r+1,r)}$ for \rf{hol} and the coefficients 
$(V,W)_k^{(r,r+1)}$ and $(W,V)_k^{(r,r+1)}$ for \rf{ahol}.

\vspace*{3mm}

\noindent
{\bf Lemma 2.}  {\it Let $\lambda^{(0)}, \rho^{(0)}\in H^0$. 
The expression 
\begin{eqnarray}
\int_{C_{\epsilon,z}}\lambda^{(1)}(w)
d\rho^{(1)}(z)-(-1)^{(n_{\rho}+1)(n_{\lambda}+1)}\int_{C_{\epsilon,z}}\rho^{(1)}(w)
d\lambda^{(1)}(z)
\end{eqnarray}
is always exact with respect to the de Rham differential.}\\
{\bf Proof.} Let's denote $\lambda^{(1)}\equiv  \lambda(z)dz-\b \lambda(z)d\b z$ and 
$\rho^{(1)}\equiv  \rho(z)dz-\b \rho(z)d\b z$. Then, showing that 
\begin{eqnarray}\label{rl}
\int_{C_{\epsilon,z}}\lambda^{(1)}(w)(\p\b \rho(z)+\bp\rho(z))-(-1)^{(n_{\rho}+1)(n_{\lambda}+1)}
\int_{C_{\epsilon,z}}\rho^{(1)}(w)(\p\b \lambda(z)+\bp \lambda(z))
\end{eqnarray}
reduces to sum $\p \b \alpha+\bp \alpha$ for some operators $\b \alpha$ and $\alpha$, 
we prove Lemma 2. 
Let's consider the first term in \rf{rl}. 
Recalling that the action of $\p \cdot$ and $\bp \cdot$ is equivalent to the action of Virasoro generators 
$[L_{-1},\cdot]$ and $[\b L_{-1},\cdot]$ correspondingly, the first term of \rf{rl} 
can be rewritten as follows:
\begin{eqnarray}
&&[L_{-1}, \int_{C_{\epsilon,z}}\lambda^{(1)}(w)\b \rho(z)]+
[\b L_{-1}, \int_{C_{\epsilon,z}}\lambda^{(1)}(w)\rho(z)]-\nonumber\\
&& \int_{C_{\epsilon,z}}([L_{-1},\lambda(w)]dw-[L_{-1},\b \lambda(w)]d\b w)
\b \rho(z)-\nonumber\\
&&\frac{1}{2\pi i}\int_{C_{\epsilon,z}}([\b L_{-1},\lambda(w)]dw-[\b L_{-1},\b \lambda(w)]d\b w)
\rho(z).
\end{eqnarray}
We can see that the first two terms in the formula above 
is represented in the needed form, while the other ones can be reexpressed:
\begin{eqnarray}\label{lr1}
&&\int_{C_{\epsilon,z}}d\b w([\b L_{-1},\lambda(w)]+[L_{-1},\b \lambda(w)])
\b \rho(z)-\nonumber\\
&&\int_{C_{\epsilon,z}}dw([\b L_{-1},\lambda(w)]+[ L_{-1},\b \lambda(w)]) \rho(z),
\end{eqnarray} 
using the fact that the integral of the total derivative vanishes. 
Let's compare this with the second term in \rf{rl}:
\begin{eqnarray}\label{lr2}
(-1)^{(n_{\rho}+1)(n_{\lambda}+1)}\int_{C_{\epsilon,z}}\b \rho(w)d\b w([L_{-1},\b \lambda](z)+[\b L_{-1}, \lambda](z))-\nonumber\\
(-1)^{(n_{\rho}+1)(n_{\lambda}+1)}\int_{C_{\epsilon,z}}\rho(w)d w([L_{-1},\b \lambda](z)+[\b L_{-1}, \lambda](z)).
\end{eqnarray} 
In order to see that the sum of \rf{lr1} and \rf{lr2} is equal to the sum $\p\b \beta +\bp \beta$ for some 
$\beta$, one needs to use Lemma 1. $\blacksquare$ \\
{\bf Proof of Proposition 2.2.} The first part easily follows from Lemma 1. Let's prove the second one. 
First of all, let's write down the expression for $[Q, K(\phi^{(0)}, \psi^{(0)})]$. Using the descent formulas 
\rf{desc}, we get:
\begin{eqnarray}
&&[Q, K(\phi^{(0)}, \psi^{(0)})]=-K( [Q,\phi^{(0)}], \psi^{(0)})-
(-1)^{n_{\phi}}K( \phi^{(0)}, [Q,\psi^{(0)}])\nonumber\\
&&+(-1)^{n_{\phi}+1}\mathcal{P}\int_{C_{\epsilon,z}}\phi^{(1)}d\psi^{(1)}+(-1)^{n_{\phi}n_{\psi}+n_{\psi}+1}
\mathcal{P}\int_{C_{\epsilon,z}}\psi^{(1)}d\phi^{(1)}.
\end{eqnarray} 
Considering the last two terms, we see that they give the exact 2-form by Lemma 2. This proves the second part of the Proposition. $\blacksquare$

\vspace*{3mm}

\noindent
{\bf Remark.} Lemma 1 and Lemma 2 are the generalizations on the case of arbitrary ghost number of 
Propositions 2.1. and 2.2. of \cite{zeit}.\\

\subsection{Generalized Maurer-Cartan equations and conformal invariance.} 
In the paper \cite{zeit}, we considered the equation of the conservation of the BRST charge in the conformal field theory perturbed 
by the operator-valued 2-form $\phi^{(2)}$. More precisely, we considered it up to the second order in the formal parameter 
(coupling constant). Namely, we expanded $\phi^{(2)}=\sum^{\infty}_{n=0}\phi_n^{(2)}t^n$ and the resulting equations up to the 
second order in $t$ were:
\begin{eqnarray}\label{cons}
&&[Q,\phi_1^{(2)}](z)=d\psi_1^{(1)}(z),\nonumber\\ 
&&[Q,\phi_2^{(2)}](z)+\frac{1}{2\pi i}\int_{C_{\epsilon, z}}\psi^{(1)}_1\phi_1^{(2)}(z)=d\psi_2^{(1)}(z), 
\end{eqnarray}
where $\psi_1^{(1)}, \psi_2^{(1)}\in H^1$ are of ghost number 1, and $\psi_1^{(1)}, \psi_2^{(1)}, \phi_2^{(2)}$ are $\epsilon$ dependent. 
Under the certain conditions, $\epsilon$-independent slice of equations \rf{cons} is shown to give the equations of 
conformal invariance at one loop in the case of two different (the first order and the second order) sigma models. 
However, one equation was missing, the so-called dilaton equation. In this paper, we fill this gap, 
i.e. we formulate the operator equations which provide the expression for the total beta-function of perturbed theory up to the second order in the
formal parameter $t$. Namely, we claim that equations should be of the following form:
\begin{eqnarray}\label{mc1}
&&[Q,\phi_1^{(0)}]=0,\\
&&\label{mc2}[Q,\phi,_2^{(0)}]+\half M(\phi_1^{(0)}, \phi_1^{(0)})=0,
\end{eqnarray}
such that $\phi_i^{(2)}=dz\wedge d\b zb_{-1}\t b_{-1}\phi_i^{(0)}$ is of ghost number 2. 
First of all, Proposition 2.2. leads to the following.

\vspace*{3mm}

\noindent
{\bf Proposition 2.3.} {\it Applying $b_{-1}\t b_{-1}$ operator to equations \rf{mc1}, \rf{mc2}, we find the following ones:
\begin{eqnarray}\label{consimp}
&&[Q,\phi_1^{(2)}]=d\phi_1^{(1)},\nonumber\\ 
&&[Q,\phi_2^{(2)}]+\frac{1}{2}K(\phi_1^{(0)}, \phi_1^{(0)})=d\chi_2^{(1)}, 
\end{eqnarray}
where $\chi_2^{(1)}\in H^1$ is some 1-form.}

\vspace*{3mm}

\noindent
{\bf Remark 1.} Equations \rf{consimp} can also be represented in the Maurer-Cartan form \cite{zeit}. In order to do this, one defines a nilpotent 
operator $D=d+\theta Q$ and superfields $\Phi_1=\phi_1^{(2)}+\theta \phi_1^{(1)}$, $\Phi_2=\phi_2^{(2)}+\theta \chi_2^{(1)}$, 
where $\theta$ is the Grassman number anticommuting with $d$. Then, defining a bilinear operation 
$\mathcal{K}(\Phi_1, \Phi_1)=\theta K(\phi_1^{(0)},\phi_1^{(0)})$,  equations 
\rf{consimp} have the following form:
\begin{eqnarray}
D\Phi_1=0, \quad D\Phi_2+ \mathcal{K}(\Phi_1, \Phi_1)=0.
\end{eqnarray}

\vspace*{3mm}

\noindent
{\bf Remark 2.} The expression $Q+\int \phi^{(1)}$ can be interpreted as a deformed BRST charge in the background of $\phi^{(2)}$, see e.g. \cite{verlinde}.

\vspace*{3mm}

\noindent
Therefore, from Proposition 2.3. we see that $\epsilon$-independent slice of equations \rf{cons} with $\psi^{(1)}=\phi^{(1)}$
can be obtained as descent from \rf{mc1}, \rf{mc2}. 
Next, we define the subspace $S^0$ of $H^0$ in which we will seek the solutions of equations \rf{mc1}, \rf{mc2}.

\vspace*{3mm}

\noindent
{\bf Definition 2.3.} {\it The space $S^0$ consists of the elements $\phi^{(0)}\in H^0$ which enjoy three properties}:\\
{\bf 1.} $n_{\phi}=2$,\\
{\bf 2.} $b^{-}_0\phi^{(0)}=0$,\\
{\bf 3.} $b_{i}\t b_{j}\phi^{(0)}=0$ {\it if $i+j>-1$, $b_{i}b_{j}\phi^{(0)}=\t b_{i}\t b_{j}\phi^{(0)}=0\ $ if $\ i+j>0$}.

\vspace*{3mm}

\noindent
{\bf Remark.} Condition 2 in  Definition 2.3. is usual in canonical SFT \cite{zwiebach}. Condition 3 is 
included to get rid of additional fields, which however usually decouple from the equations on $V$, obtained from \rf{mc1}, \rf{mc2}.
 
\vspace*{3mm}

\noindent
As we see, the general form of the element from $\phi^{(0)}\in S^0$ is as follows:
\begin{eqnarray}
\phi^{(0)}=\t cc V+c(\p c +\bp \t c)W- \t c  (\p c +\bp \t c)\b W+1/2c\p^2 c U-1/2\t c\bp^2 \t c\b U.
\end{eqnarray}
Here, $V$ is a perturbation operator, and we will refer to $W$, $\b W$ as $gauge$ terms and $U$, $\b U$ as $dilatonic$ terms.
We will keep this notation in the following. 

In order to get in touch with the examples, we consider the following assumptions related to perturbation 2-form 
$\phi^{(2)}$.

\vspace*{3mm}

\noindent
{\bf Assumptions.} {\it Let perturbation 2-form be 
$\phi^{(2)}=dz\wedge d\b z V(z, \b z)$, where the perturbation operator $V\in H^0$ of ghost number 0. 
We will consider the perturbation operators which satisfy two conditions:}\\
1. {\it $L_mV=\b L_nV=0$ for $m,n>1$ and $(L_0V)=(\b L_0V)$, where $L_m$ and $\b L_n$ are the corresponding Virasoro generators.}\\
2. {\it The operator product coefficients $(V,V)_l^{(m,n)}=0$ for $m>2$ or $n>2$.}

\vspace*{3mm}

\noindent
{\bf Remark 1.} The assumptions above correspond to two examples we already  considered in the context of equations \rf{cons} 
in \cite{zeit}, \cite{zeit2}.

\vspace*{3mm}

\noindent
{\bf Remark 2.} The condition from point 1 of Assumptions can be rewritten by means of the BRST operator and $b, \t b$ -ghosts in the following way: 
\begin{eqnarray}
b_i[Q, \phi^{(2)}]=\t b_i[Q, \phi^{(2)}]=0,
\end{eqnarray}
when $i>1$.

\vspace*{3mm}

\noindent
{\bf Proposition 2.4.} {\it Let $\phi_i^{(0)}$ (i=1,2) be the elements of $S^0$ such that $\phi_i^{(2)}=dz\wedge d\b zV_i(z, \b z)$ and $V_i$ satisfy Assumptions above. Then  equation \rf{mc1} leads to the operator equations on $V_1$:
\begin{eqnarray}\label{comp1}
&&(L_0V_1)(z)-V_1(z)+L_{-1}W_1+ L_{-1} W_1=0, \nonumber\\
&&W_1=-1/2((\b L_1 V_1)+L_{-1}\b U_1), \quad \b W_1=-1/2((L_1 V_1)+\b L_{-1}U_1),\\
&&\label{dil1}L_1W_1=0, \quad \b L_1\b W_1=0,
\end{eqnarray}
and equation \rf{mc2} leads to the operator equations on $V_2$: 
\begin{eqnarray}\label{comp2}
&&(L_0V_2)-V_2-1/2(V_1,V_1)_0^{(1,1)}+1/2
( \b W_1,V_1)_0^{(0,1)}-\nonumber\\
&&1/2
( V_1,\b W_1)_0^{(0,1)}
+1/2(W_1,V_1)_0^{(1,0)}-\nonumber\\
&&1/2(V_1,W_1)_0^{(1,0)}+\b L_{-1}W_2+ L_{-1} \b W_2=0,
\end{eqnarray}
\begin{eqnarray}
&&\label{W}\b W_2=-1/2((L_1 V_2)-(V_1,V_1)_0^{(2,1)}+(\b W_1,V_1)_0^{(1,1)}\nonumber\\
&&+(W_1,V_1)_0^{(2,0)}+1/2(U_1,V_1)_0^{(1,0)}-
1/2(V_1,U_1)_0^{(1,0)}+\b L_{-1}U_2),\nonumber\\
&&W_2=-1/2((\b L_1 V_2)-(V_1,V_1)_0^{(1,2)}+(W_1,V_1)_0^{(1,1)}+\nonumber\\
&&(\b W_1,V_1)_0^{(0,2)}+1/2(\b U_1,V_1)_0^{(0,1)}-
1/2(V_1,\b U_1)_0^{(0,1)}+L_{-1}\b U_2),
\end{eqnarray}
\begin{eqnarray}
&&\label{dil2}2L_1W_2-2L_0U_2+(U_1,W_1)_0^{(1,0)}-(W_1,U_1)_0^{(1,0)}+2(W_1,W_1)_0^{(2,0)}\\
&&-(V_1,W_1)_0^{(2,1)}+(V_1,U_1)_0^{(1,1)}+2(\b W_1,W_1)_0^{(1,1)}-(\b W_1,U_1)_0^{(0,1)}=0\nonumber\\
&&\label{dil22}2(\b L_1\b W_2)-2\b L_0\b U_2+(\b U_1,\b W_1)_0^{(0,1)}-(\b W_1,\b U_1)_0^{(0,1)}+2(\b W_1,\b W_1)_0^{(0,2)}\\
&&-(V_1,\b W_1)_0^{(1,2)}+(V_1,\b U_1)_0^{(1,1)}+2(W_1,\b W_1)_0^{(1,1)}-(W_1,\b U_1)_0^{(0,1)}=0\nonumber.
\end{eqnarray}}

\vspace*{3mm}

\noindent
The Proof can be obtained by the direct calculation.\\
In the next subsection, we will consider an example of perturbed 2d conformal field theory, familiar from \cite{zeit}, and observe that 
the corresponding equations of conformal invariance obtained by appropriate renormalization techniques coincide with \rf{comp1}-\rf{dil2}. 

One of the important features 
of equations \rf{mc1}, \rf{mc2} and, therefore, of \rf{comp1}-\rf{dil2} is that they are automatically covariant, that is they are invariant under 
symmetry transformations. 
Really, it is easy to see that due to Proposition 2.1. the following statement holds.

\vspace*{3mm}

\noindent
{\bf Proposition 2.5.} {\it Let $\phi_1^{(0)}$, $\phi_2^{(0)}\in H^0$ be of ghost number 2. 
Equations \rf{mc1}, \rf{mc2} are invariant under the following symmetry transformations:
\begin{eqnarray}\label{transf}
\delta\phi_1^{(0)}=\varepsilon[Q,\xi_1^{(0)}], \quad 
\delta\phi_2^{(0)}=\varepsilon([Q,\xi_2^{(0)}]+M(\xi_1^{(0)}, \phi_1^{(0)})),
\end{eqnarray}
where $\xi_{1,2}^{(0)}\in H^0$ are of ghost number 1 and $\epsilon$ is infinitesimal.}

\vspace*{3mm}

\noindent
{\bf Remark.} We also note that in the next subsection, we will meet an example in which $\xi_{2}^{(0)}$ depends on $\xi_1^{(0)}$ and 
$\phi_1^{(0)}$, generating the structure of algebroid to the transformations \rf{transf}. 

\subsection{Example: closed strings in background fields. }
Let's consider the theory of D free massless bosons with the action: 
\begin{eqnarray} 
S=\frac{1}{4\pi\alpha'}\int d^2 z \eta_{\mu\nu}\p X^{\mu}\bp X^{\nu},
\end{eqnarray}
where  $\eta_{\mu\nu}$ is a constant nondegenerate symmetric matrix, 
$\mu,\nu=1,...,D$ and $d^2z=idz\wedge d\b z$.\\
\hspace*{5mm}The operator product, generated by the free boson field theory, is as follows:
\begin{eqnarray}
X^{\alpha}(z_1)X^{\beta}(z_2)\sim -\eta^{\alpha\beta}\alpha'\log|(z_{1}-z_2)/\mu|^2,
\end{eqnarray}
where $\mu$ is some nonzero parameter. 
The energy-momentum tensor is given by such an expression: 
\begin{eqnarray}
T=-(2\alpha')^{-1}\p X^{\mu}\p X_{\mu}, \quad 
\t T=-(2\alpha')^{-1}\bp X^{\mu}\bp X_{\mu}.
\end{eqnarray}
Let's consider the sigma-model action, which describes strings moving in the background metric $G_{\mu\nu}$ and a dilaton $\Phi$. It is 
written as follows:
\begin{eqnarray} \label{sigma}
S=\int_{\Sigma} d^2 z (\frac{1}{4\pi\alpha'}G_{\mu\nu}(X)\p X^{\mu}\bp X^{\nu}+\frac{1}{2}\sqrt{\gamma}R^{(2)}(\gamma)\Phi(X)),
\end{eqnarray}
where $R^{(2)}$ is a curvature on a Riemann surface $\Sigma$. 

We also assume that 
$G_{\mu\nu}(X), \Phi(X)$ are expanded with respect to some formal parameter t:
\begin{eqnarray}\label{gexp} 
&&G_{\mu\nu}=\eta_{\mu\nu}-th_{\mu\nu}(X)-t^2s_{\mu\nu}(X)+O(t^3),\\
&&\Phi=t\Phi_1+t^2\Phi_2+O(t^3),
\end{eqnarray}
where $\eta_{\mu\nu}$ is independent of $X$. 
This allows dealing with 
$\phi^{(2)}=(2\alpha')^{-1}(\eta_{\mu\nu}-G_{\mu\nu})\p X^{\mu}\bp X^{\nu}dz\wedge d\b z$ as 
a perturbation 2-form and applying the Maurer-Cartan equations to this case. 
But as we explained in \cite{zeit}, we miss some terms. The reason is that our formalism does not allow to 
take into account the so-called contact terms from perturbation theory, namely those, which contain 
$\delta$-functions in operator products. Therefore, they should be added to the action. 
In \cite{zeit}, we explicitly constructed these contact terms at the second order of the perturbation theory 
and calculated the proper coefficients for them to enter the action. 
So, we have to consider the following perturbation 2-forms:
\begin{eqnarray}\label{gravpert}
&&\phi^{(2)}_1=dz\wedge d\b z(2\alpha')^{-1}h_{\mu\nu}(X)\p X^{\mu}\bp X^{\nu},\\
&&\phi^{(2)}_2=dz\wedge d\b z(2\alpha')^{-1}(s_{\mu\nu}(X)\p X^{\mu}\bp X^{\nu}+
1/2h_{\mu\rho}(X)\eta^{\rho\sigma}h_{\nu\sigma}(X)\p X^{\mu}\bp X^{\nu})\nonumber,
\end{eqnarray}
where we included an additional $bivertex$ operator (which is a contribution of contact terms) in $\phi^{(2)}_2$. 
In this case, the following proposition holds.

\vspace*{3mm}

\noindent
{\bf Proposition 2.6.} 
{\it Constraints \rf{comp1}-\rf{dil2} for \rf{gravpert}, where  
$U_i\equiv U_i(X)$ and $\b U_i\equiv \b U_i(X)$, lead to the 
Einstein equations  
\begin{eqnarray}\label{einst}
&&R_{\mu\nu}+2\nabla_{\mu}\nabla_{\nu}\Phi=0,\nonumber\\
&&R+4\nabla_{\mu}\nabla^{\mu}\Phi-4\nabla^{\mu}\Phi\nabla_{\mu}\Phi=0
\end{eqnarray}
expanded up to the second order in t, where the expansion of metric and dilaton is given by formulas 
\rf{gexp}, such that correspondence between dilaton and $U,\b U$-variables is given by the formula
\begin{eqnarray}
&&\Phi_1=1/2t(U_1+\b U_1-1/2h),\nonumber\\
&&\Phi_2=1/2(U_2+\b U_2-1/2s-1/4h_{\mu\nu}h^{\mu\nu}),
\end{eqnarray}
where $h=\eta^{\mu\nu}h_{\mu\nu}$ and $s=\eta^{\mu\nu}s_{\mu\nu}$.}

\vspace*{3mm}

\noindent
The Proof is given in  Appendix.\\
Thus we see,  equations \rf{einst}, corresponding to 1-loop conformal invariance conditions for the sigma model \rf{sigma}, 
up to the second order of expansion in the formal parameter $t$ have the generalized Maurer-Cartan structure given by 
equations \rf{mc1}, \rf{mc2}. 

In the end of subsection 2.2, we mentioned that equations \rf{mc1}, \rf{mc2} possess symmetries accurately described in Proposition 2.5. 
Let's look how it works in this case. Action \rf{sigma} and equations \rf{einst} are invariant under the diffeomorphism transformations. 
The infinitesimal change of metric tensor is as follows:
\begin{eqnarray}\label{gt}
G_{\mu\nu}\to G_{\mu\nu}-\varepsilon(\nabla_{\mu}v_{\nu}+\nabla_{\nu}v_{\mu}),
\end{eqnarray}   
where $\varepsilon$ is infinitesimal. Let's expand
\begin{eqnarray}\label{vexp}
v_{\nu}=tv^1_{\nu}+t^2v^2_{\nu}+O(t^3).
\end{eqnarray}   
Therefore, at the first order in $t$, the transformation is given by:
\begin{eqnarray}\label{sym1}
h_{\mu\nu}\to h_{\mu\nu}+\varepsilon(\p_{\mu}v_{\nu}+\p_{\nu}v_{\mu}).
\end{eqnarray}  
Let's consider the following operators of ghost number 1:
 \begin{eqnarray}\label{xi1}
\xi_1^{(0)}=(2\alpha')^{-1}(cv^1_{\mu}(X)\p X^{\mu} - \tc v^1_{\mu}(X)\b \p X^{\mu}d\b z).
\end{eqnarray} 
It is easy to see that the transformation
\begin{eqnarray}\label{transf1}
\phi_1^{(0)}\to \phi_1^{(0)}+\varepsilon[Q, \xi_1^{(0)}],
\end{eqnarray} 
where $\phi_1^{(0)}$ is as in Proposition 2.6., reproduces \rf{sym1}. At the second order the situation is more complicated:
\begin{eqnarray}\label{deltas}
&&s_{\mu\nu}\to s_{\mu\nu}+\varepsilon(\p_{\mu}v^2_{\nu}+\p_{\nu}v^2_{\mu}-2\Gamma^{\rho}_{\mu\nu}v^1_{\rho}=\nonumber\\
&&\p_{\mu}v^2_{\nu}+\p_{\nu}v^2_{\mu}+v^1_{\rho}\eta^{\rho\sigma}(-\p_{\sigma}h_{\mu\nu}+
\p_{\mu}h_{\sigma\nu}+\p_{\nu}h_{\sigma\mu})).
\end{eqnarray} 
One might think that changing the indices from 1 to 2 in \rf{xi1}, one gets the expression for $\xi_2^{(0)}$ which will reproduce the 
diffeomorphism transformation \rf{deltas}. However, the situation appears to be more complicated: the expression for $v^2_{\nu}$ should be 
improved by the terms ${v^{1}}^{\mu}h_{\mu\nu}$, the emergence of which can be substantiated by the same reason as the bivertex 
operator appeared in $\phi^{(2)}_2$. By straightforward calculation, one can obtain that 
\begin{eqnarray}\label{xi2}
\xi_2^{(0)}=(2\alpha')^{-1}(c(v^2_{\mu}+3/4{v^{1}}^{\nu}h_{\nu\mu})\p X^{\mu} - \tc (v^2_{\mu}(X)+3/4{v^{1}}^{\nu}h_{\nu\mu})\b \p X^{\mu}d\b z)
\end{eqnarray} 
together with \rf{xi1} by means of the formula 
\begin{eqnarray}\label{transf2}
\phi_2^{(0)}\to \phi_2^{(0)}+\varepsilon([Q,\xi_2^{(0)}]+M(\xi_1^{(0)}, \phi_1^{(0)}))
\end{eqnarray}
reproduces transformation \rf{deltas} modulo the terms of higher order in $\alpha'$. 
It should be noted that we already met such additional terms during the study of 
the symmetries of the equation describing the conservation of BRST current \cite{zeit}.
 
Let's now summarize the results concerning symmetries in the proposition.  

\vspace*{3mm}

\noindent
{\bf Proposition 2.7.} {\it The transformations \rf{transf1}, \rf{transf2}, where $\xi_1^{(0)},\xi_2^{(0)}$ are given by \rf{xi1}, \rf{xi2}, and 
perturbation operators  are given by \rf{gravpert}, reproduce the infinitesimal diffeomorphism transformations expanded up to the second order in 
the formal parameter modulo the terms of higher order in $\alpha'$.}

\vspace{3mm}

\noindent{\bf Remark 1.} The similar results, namely the reproduction of the conformal invariance conditions and 
their symmetries from  equations \rf{mc1}, \rf{mc2}, 
were obtained in the case of the perturbed beta-gamma systems \cite{nekrasov}, \cite{lmz} in \cite{zeit2}. 
One of the differences which is worth mentioning 
is that the equations of conformal invariance at one loop in that model 
appear to be bilinear in the perturbation operator, and therefore the second order approximation appears to be exact.

\vspace{3mm}

\noindent{\bf Remark 2.} In \cite{sengh}, the nonlinear corrections to the symmetries of linearized Einstein equations were obtained 
in the context of SFT.

\section{Open Strings and Yang-Mills Equations}

{\bf 1. Notation and Conventions.} 
Throughout this section, we will deal with an example of boundary conformal field theory, i.e. the open string on a disc (conformally mapped to the upper half-plane), see e.g. \cite{pol}. 
Namely, we will consider the theory with $D$ scalar bosons, such that the operator products between scalar fields are: 
\begin{eqnarray}
X^{\mu}(z_1)X^{\nu}(z_2)\sim -\eta^{\mu\nu}\alpha'\log|(z_1-z_2)/\mu|^2-\eta^{\mu\nu}\alpha'\log|(z_1-\b z_2)/\mu|^2,
\end{eqnarray}
where $\eta^{\mu\nu}$ is the constant metric in the flat $D$-dimensional space either of Euclidean or Minkovski signature. In this theory,
the operators  have the following operator products on the real line:
\begin{eqnarray}\label{ope2}
V(t_1)W(t_2)\sim \sum^n_{k=-\infty}(t_1-t_2)^{-k}(V,W)_l^{(k)}(\log|(t_1-t_2)/\mu|^2)^k
\end{eqnarray}
for some $n$. We also introduce the energy momentum tensor:
\begin{eqnarray}
T=-\frac{1}{2\alpha'}\p X^{\mu}\bp X_{\mu}
\end{eqnarray}
and associated BRST operator:
\begin{eqnarray}\label{brsto}
Q=\oint dz (cT+bc\p c), 
\end{eqnarray}
where the operator products between ghost fields are as usual $c(z)b(w)\sim \frac{1}{z-w}$. The same way,
 we define the ghost number operator:
\begin{eqnarray}
N_g=-\oint dz bc.
\end{eqnarray}
We also introduce the space $F^0$ of differential polynomials in $c$-ghost field, where the coefficients are matter field operators. This space is obviously graded 
with respect to the ghost number operator. 
For any $\phi^{(0)}\in F^0$, which is an eigenvector of $N_g$, we will denote the corresponding eigenvalue by $n_{\phi}$, 
i.e. ghost number. As in section 2, we define the space of operator valued 1-forms $\phi^{(1)}=Vdz$, where $V\in F^1$ 
with associated equation: for any given $\phi^{(0)}\in F^0$ one can define $\phi^{(1)}\in F^1$ such that 
\begin{eqnarray}
[Q,\phi^{(1)}]=d\phi^{(0)}-[Q,\phi^{(0)}]^{(1)}.
\end{eqnarray}
The main characters of this section will be the elements of the tensor product $F^0_{\mathbf{g}}=F^0\otimes \mathbf{g}$, 
where $\mathbf{g}$ is some Lie algebra. 

\vspace*{3mm}

\noindent
\noindent{\bf 2. Generalized Maurer-Cartan structures and Yang-Mills equations.}\\
Let's consider two operators $\phi^{(0)}(t),\psi^{(0)}(t)\in F_{\mathbf{g}}^0$. Then, the expression
\begin{eqnarray}
[\phi^{(0)}(t+\epsilon),\psi^{(0)}(t)],
\end{eqnarray}
where $t$ lies on the real axis and $[,]$ means the commutator in Lie algebra $\mathbf g$. 
Due to \rf{ope2}, this object is  the series in $\epsilon$ and $\log(\epsilon/\mu)$, therefore, this allows us to define 
the following operation.

\vspace*{3mm}

\noindent
{\bf Definition 4.1.} {\it For any two operators $\phi^{(0)}(t),\psi^{(0)}(t)\in F_{\mathbf{g}}^0$ we define 
a bilinear operation $R: F_{\mathbf{g}}\otimes F_{\mathbf{g}}\to F_{\mathbf{g}}\ $ :
\begin{eqnarray}
R(\phi^{(0)},\psi^{(0)})(t)=\mathcal{P}[\phi^{(0)}(t+\epsilon),\psi^{(0)}(t)]-(-1)^{n_\phi n_{\psi}}
\mathcal{P}[\psi^{(0)}(t+\epsilon),\phi^{(0)}(t)],
\end{eqnarray}
where $\mathcal{P}$ is the projection on the $\epsilon^0(\log(\epsilon/\mu))^0$ term and t lies on the real axis.}

\vspace*{3mm}

\noindent
This operation satisfies the property which is very similar to that from Proposition 2.1.

\vspace*{3mm}

\noindent
{\bf Proposition 3.1.} {\it Let $\phi^{(0)}(t),\psi^{(0)}(t)\in F_{\mathbf{g}}^0$. Then 
\begin{eqnarray}
[Q,R(\phi^{(0)},\psi^{(0)})]=R([Q,\phi^{(0)}],\psi^{(0)})+(-1)^{n_{\phi}}R(\phi^{(0)},[Q,\psi^{(0)}]),
\end{eqnarray}
where Q is BRST operator \rf{brsto}.}

\vspace*{3mm}

\noindent
The proof directly follows from the definition. 
Now, since we got the bilinear operation, we are able to construct the second order approximation to the generalized
 Maurer-Cartan equation: 
\begin{eqnarray}
[Q,\phi^{(0)}]+\frac{1}{2}R(\phi^{(0)},\phi^{(0)} )+...
\end{eqnarray}
like we did in the previous section, 
i.e. we expand $\phi^{(0)}=\sum^{\infty}_{n=1}t^n\phi_n^{(0)}$ by means of the formal parameter $t$ and 
consider the equations which emerge in the first and the second order:
\begin{eqnarray}\label{mco}
[Q,\phi_1^{(0)}]=0, \quad [Q,\phi_2^{(0)}]+\frac{1}{2}R(\phi_1^{(0)},\phi_1^{(0)} )=0.
\end{eqnarray}
In order to get in touch with Yang-Mills theory, we need to put some conditions on $\phi^{(0)}$. 
Namely, we will plug operators $\phi^{(0)}\in F^0_{\mathbf{g}}$, which satisfy the following conditions:  
\begin{eqnarray}\label{cond}
n_{\phi}=1,\quad [b_{-1},\phi^{(0)}]=A_{\mu}(X)\p X^{\mu},\quad [b_i, \phi^{(0)}]=0 \quad (i>0),
\end{eqnarray}
in  equations \rf{mco}. 
Here, $A_{\mu}$ are the components of a Lie algebra-valued 1-form and normal ordering is implicit. Then, the following statement holds.

\vspace*{3mm}

\noindent
{\bf Proposition 3.2.} {\it Let's consider $\phi^{(0)}$ satisfying  conditions \rf{cond}. Then, equations \rf{mco} at the first order in 
$\alpha'$ are equivalent to the following equations:
\begin{eqnarray}\label{ym}
&&\p_{\mu}\p^{\mu}A^1_{\nu}-\p_{\nu}\p^{\mu}A^1_{\mu}=0,\\
&&\p_{\mu}\p^{\mu}A^2_{\nu}-\p_{\nu}\p^{\mu}A^2_{\mu}+[\p^{\mu}A^1_{\mu},A^1_{\nu}]+2[A^1_{\mu}, \p^{\mu}{A^1}_{\nu}]-
[A^1_{\mu},\p_{\nu}{A^1}^{\mu}]=0\nonumber,
\end{eqnarray}
where $[b_{-1},\phi_i^{(0)}]=A^i_{\mu}\p X^{\mu}$ (i=1,2)
and the indices are raised and lowered with respect to the metric $\eta^{\mu\nu}$.}\\
{\bf Proof.} From  conditions \rf{cond} we find that 
\begin{eqnarray}
\phi_i^{(0)}=cA^i_{\mu}\p X^{\mu}-\p cW_i,
\end{eqnarray}
where $W_i$ are some ``matter'' operators. Let's consider the coefficient of $c\p^2 c$ in the expression 
$[Q,\phi_1^{(0)}]$. It is easy to see that it is equal to $W_1-\alpha'\p^{\mu}A^1_{\mu}$. 
Therefore, 
\begin{eqnarray}
W_1=\alpha'\p^{\mu}A^1_{\mu}.
\end{eqnarray}
The only term which is left in $[Q,\phi_1^{(0)}]$ is that, proportional to $c\p c$, such that the proportionality coefficient is 
$\alpha'(2\p_{\mu}\p^{\mu}A^1_{\nu}-2\p_{\nu}\p^{\mu}A^1_{\mu})\p X^{\nu}$. Therefore, the following equation holds:
\begin{eqnarray}
\p_{\mu}\p^{\mu}A^1_{\nu}-\p_{\nu}\p^{\mu}A^1_{\mu}=0.
\end{eqnarray}
Thus, we proved the first part. To prove the second part, we first notice that 
\begin{eqnarray}
&&R(\phi_1^{(0)},\phi_1^{(0)})=2\mathcal{P}[\phi_1^{(0)}(t+\epsilon),\phi_1^{(0)}(t)]=\nonumber\\
&&\alpha'c\p c(2[\p^{\mu}A^1_{\mu},A^1_{\nu}]+4[A^1_{\mu}, \p^{\mu}{A^1}_{\nu}]-
2[A^1_{\mu},\p^{\nu}{A^1}^{\mu}]+O(\alpha'^2)).
\end{eqnarray}
Therefore, remembering lessons of the proof of the first part, we find that $W_2=\alpha'\p^{\mu}A^2_{\mu}$ and, therefore, the second 
of 
equations \rf{mco} at the order $\alpha'$ gives the following equation:
\begin{eqnarray}
\p_{\mu}\p^{\mu}A^2_{\nu}-\p_{\nu}\p^{\mu}A^2_{\mu}+[\p^{\mu}A^1_{\mu},A^1_{\nu}]+2[A^1_{\mu}, \p^{\mu}{A^1}_{\nu}]-
[A^1_{\mu},\p_{\nu}{A^1}^{\mu}]=0.
\end{eqnarray}
This finishes the proof. $\blacksquare$

\vspace*{3mm}

\noindent
One can easily notice that equations \rf{ym} coincide with the Yang-Mills equations
\begin{eqnarray}
\p_{\mu}F^{\mu\nu}+[A_{\mu},F^{\mu\nu}]=0,\quad F_{\mu\nu}=\p_{\mu}A_{\nu}-\p_{\nu}A_{\mu}+[A_{\mu}, A_{\nu}]
\end{eqnarray}
expanded up to the second order in the formal parameter $t$, such that the expansion of the gauge field $A_{\mu}$ is as follows: 
$A_{\mu}=tA^1_{\mu}+t^2A^2_{\mu}+O(t^3)$. 

\vspace{3mm}

\noindent{\bf Remark.} It is worth noting that in  papers 
\cite{taylor}, \cite{berkovits}, the Yang-Mills action was obtained from effective action of open SFT and WZW-like open 
super SFT correspondingly. 

\vspace{3mm}

\noindent
The next step is to figure out how the gauge symmetries appear in this formalism. The following proposition will help in this direction:

\vspace*{3mm}

\noindent
{\bf Proposition 3.3.} {\it Let $\phi_1^{(0)}, \phi_2^{(0)}\in F_{\mathbf{g}}^{(0)}$ be of ghost number 1. 
Then, the following transformations generate the symmetries of 
equations \rf{mco}: 
\begin{eqnarray}\label{sym}
\phi_1^{(0)}\to \phi_1^{(0)}+\epsilon[Q,\lambda_1^{(0)}], \quad    
\phi_2^{(0)}\to \phi_2^{(0)}+\epsilon([Q,\lambda_2^{(0)}] +R(\phi_1^{(0)},\lambda_1^{(0)})),
\end{eqnarray}
where $\lambda_i^{(0)}\in F_{\mathbf{g}}^{(0)}$ are of ghost number 0 (i=1,2).}

\vspace*{3mm}

\noindent
The proof directly follows  from  Proposition 3.1.\\
If we consider $\lambda_i^{(0)}=\lambda_i(X)$, then  transformations \rf{sym} have the following form:
\begin{eqnarray}
&&\phi_1^{(0)}\to \phi_1^{(0)}+2c\delta A^1_{\mu}\p X^{\mu}-\alpha'\p c\p^{\mu}\delta A^1_{\mu}, \nonumber\\
&&\phi_2^{(0)}\to \phi_2^{(0)}+2c(\delta A^2_{\mu}\p X^{\mu}+O(\alpha'))-\alpha'\p c(\p^{\mu}\delta A^2_{\mu}+O(\alpha')),
\end{eqnarray}
where $\delta A^1_{\mu}=\epsilon\p_{\mu}\lambda_1$, $\delta A^2_{\mu}=\epsilon(\p_{\mu}\lambda_2+[A^1_{\mu},\lambda_1])$, which coincide with the usual Yang-Mills 
gauge transformations with the element of gauge transformation expanded up to the second order in $t$: $\lambda=t\lambda_1+t^2\lambda_2+O(t^3)$.
\section{Conclusion and Final Remarks}

In this paper, we have considered the formal Maurer-Cartan equations \rf{mc1} and \rf{mc2} and have shown that they lead to 
the second order approximations to the corresponding classical 
field equations, namely Einstein and Yang-Mills ones. However, our constructions 
involve the further corrections in $\alpha'$ parameter how it usually happens with beta-functions. 

Here, we make a claim that at list 
in case of perturbation by a gauge field, it is possible to redefine the operation $R$, i.e. make a restriction of it to some 
subspace, denoting the result as $R_1$, and define another graded 3-linear operation $R_2$, which together satisfy the relations 
of a homotopy Lie algebra, such that the Yang-Mills equation will be written in the form of the generalized Maurer-Cartan equation 
\begin{eqnarray}
[Q, \phi^{(0)}]+\frac{1}{2!}R_1(\phi^{(0)}, \phi^{(0)})+\frac{1}{3!}R_2(\phi^{(0)},\phi^{(0)}, \phi^{(0)})=0.
\end{eqnarray}
This subject will be studied in \cite{zeit3}. 

In the case of Einstein equations, we suggest that such redefinition can be made, however 
in contrast to Yang-Mills, due to the strong nonlinearity one might expect that a number of operations $M_n$ in the GMC equation 
should be infinite. In this respect, we note that the first order formulation of string theory in background of metric B-field 
and dilaton \cite{zeit2}, \cite{lmz} looks more promising since this formalism does not destroy the geometry and as it was shown 
in \cite{zeit2} probably will lead to the generalizations of the homotopy algebra of Courant/Dorfman brackets. 

\section*{Acknowledgements}
The author is grateful to A.S. Losev for 
introduction in the subject and fruitful discussions. 
It is important to mention that the hypotheses concerning the using of generalized Maurer-Cartan equations and $L_{\infty}$-structures  
in the context of the study of the conditions of conformal invariance belong to A.S. Losev.
The author is very grateful to I.B. Frenkel, M. Kapranov and G. Zuckerman for numerous discussions on the subject 
and to I.B. Frenkel and N.Yu. Reshetikhin for their permanent encouragement and support.

\section*{Appendix}
\addcontentsline{toc}{section}{Appendix}
{\bf Proposition 2.6.} 
{\it Constraints \rf{comp1}-\rf{dil2} for \rf{gravpert}, where  
$U_i\equiv U_i(X)$ and $\b U_i\equiv \b U_i(X)$ lead to the 
Einstein equations  
\begin{eqnarray}\label{einst2}
&&R_{\mu\nu}+2\nabla_{\mu}\nabla_{\nu}\Phi=0,\nonumber\\
&&R+4\nabla_{\mu}\nabla^{\mu}\Phi-4\nabla^{\mu}\Phi\nabla_{\mu}\Phi=0
\end{eqnarray}
expanded up to the second order in t, where the expansion of metric and dilaton is given by formulas 
\rf{gexp}, such that correspondence between dilaton and $U,\b U$-variables is given by the formula
\begin{eqnarray}
&&\Phi_1=1/2t(U_1+\b U_1-1/2h),\nonumber\\
&&\Phi_2=1/2(U_2+\b U_2-1/2s-1/4h_{\mu\nu}h^{\mu\nu}),
\end{eqnarray}
where $h=\eta^{\mu\nu}h_{\mu\nu}$ and $s=\eta^{\mu\nu}s_{\mu\nu}$.}\\
{\bf Proof.} Here, we will give the expression for the Einstein equations \rf{einst2}
with 
the metric and a dilaton expanded to the second order of perturbation parameter $t$:
\begin{eqnarray}
&&G_{\mu\nu}=\eta_{\mu\nu}-th_{\mu\nu}(X)-t^2s_{\mu\nu}(X)+O(t^3),\nonumber\\ 
&&\Phi= \Phi_0+t\Phi_1(X)+
t^2 \Phi_2(X)+O(t^3).
\end{eqnarray} 
At the first order in $t$, we have:
\begin{eqnarray}\label{h}
&&1/2\Delta h_{\mu\nu}-1/2\p_{\mu}\p_{\rho}h^{\rho}_{\nu}-1/2\p_{\nu}\p_{\rho}h^{\rho}_{\mu}+
1/2\p_{\mu}\p_{\nu}h+2\p_{\mu}\p_{\nu}\Phi_1=0,\\
&&\label{phi1}\Delta h-\p^{\mu}\p^{\nu}h_{\mu\nu}+4\p_{\mu}\p^{\nu}\Phi_1=0, 
\end{eqnarray} 
where $h=\eta^{\rho\sigma}h_{\rho\sigma}$ and $\Delta=\p_{\mu}\p^{\mu}$. The indices are raised and lowered  
by means of the flat metric $\eta^{\rho\sigma}$. 
The next order gives:
\begin{eqnarray}\label{s}
&&1/2\Delta s_{\mu\nu}-1/2\p_{\nu}\p^{\beta}s_{\beta\mu}-1/2\p_{\mu}\p^{\beta}
s_{\beta\nu}+\p_{\nu}\p_{\mu}(1/2s+2\Phi_2+1/8 h^{\rho\sigma}h_{\rho\sigma})\nonumber\\
&&+1/2(\p^{\beta}h_{\beta\xi}-\p_{\xi} (1/2h+2\Phi_1))
\eta^{\xi\rho}(\p_{\rho}h_{\nu\mu}-\p_{\mu}h_{\nu\rho}-\p_{\nu}h_{\mu\rho})\nonumber\\
&&+1/2\eta^{\xi\rho}\p_{\xi}h_{\nu\lambda}\eta^{\lambda\alpha}\p_{\rho}h_{\alpha\mu}+
1/2\eta^{\xi\rho}\eta^{\sigma\alpha}h_{\rho\sigma}\p_{\xi}\p_{\alpha}h_{\mu\nu}\nonumber\\
&&-1/2\eta^{\xi\rho}\eta^{\lambda\alpha}\p_{\lambda}h_{\xi\nu}\p_{\rho}h_{\alpha\mu}-
1/2\p_{\sigma}\p_{\mu}h_{\nu\chi}h_{\xi\alpha}\eta^{\xi\chi}\eta^{\sigma\alpha}-\nonumber\\
&&1/2\p_{\sigma}\p_{\nu}h_{\mu\chi}h_{\xi\alpha}\eta^{\xi\chi}\eta^{\sigma\alpha}+
1/4h_{\alpha\rho}\p_{\nu}\p_{\mu}h_{\xi\lambda}\eta^{\alpha\xi}\eta^{\rho\lambda}=0.\\
&&\label{phi2}\Delta s -\p^{\mu}\p^{\nu}s_{\mu\nu}+4\p_{\nu}\p^{\nu}\Phi_2+3/4\p_{\alpha}h_{\mu\nu}\p^{\alpha}h^{\mu\nu}-\nonumber\\
&&1/2\p_{\mu}h_{\nu\alpha}\p^{\nu}h_{\mu\alpha}+(\p^{\beta}h_{\beta\xi}-\p_{\xi}(1/2h+2\Phi_1))
(\p^{\xi}(1/2h+2\Phi_1)-\nonumber\\
&&\p_{\rho}h^{\rho\xi})=0.
\end{eqnarray}
Let's obtain \rf{h}, \rf{phi1} from \rf{comp1},\rf{dil1}. 
We have:
\begin{eqnarray}
&&V_1=1/2\alpha'^{-1}h_{\mu\nu}(X)\p X^{\mu}\bp X^{\nu},\\
&&V_2=1/2\alpha'^{-1}(s_{\mu\nu}(X)+
1/2h_{\mu\rho}\eta^{\rho\sigma}h_{\nu\sigma}(X))\p X^{\mu}\bp X^{\nu}.
\end{eqnarray} 
So, we just need to substitute these operators in  equations \rf{comp1}, \rf{dil1}. 
Starting from the first one
\begin{eqnarray}\label{fst}
&&(L_0V_1)-V_1-1/2L_{-1}L_1V_1-1/2\b L_{-1}\b L_1V_1-\nonumber\\
&&1/2L_{-1}\b L_{-1} (U_1+\b U_1)=0,
\end{eqnarray} 
we see that 
\begin{eqnarray}
&&(L_0V_1)-V_1=-1/4\Delta h_{\mu\nu}(X)\p X^{\mu}\bp X^{\nu},\\
&&1/2(L_1V_1+\b L_{-1}U_1)=-1/4(\p^{\beta}h_{\beta\xi}-2\p_{\xi}U_1)\bp X^{\xi},\\
&&1/2(\b L_1V_1+L_{-1} \b U_1)=-1/4(\p^{\beta}h_{\beta\xi}-2\p_{\xi}\b U_1)\p X^{\xi}.
\end{eqnarray} 
In such a way, we see that equation \rf{fst} coincides with \rf{h} if
\begin{eqnarray}
U_1+\b U_1=1/2h+2\Phi_1, 
\end{eqnarray} 
so our choice for $U$-terms was correct. Similarly, one obtains that two other equations:
\begin{eqnarray}
L_1W_1=0,\qquad \b L_1\b W_1=0
\end{eqnarray} 
coincide with \rf{dil1} since
\begin{eqnarray}
L_1W_1=\b L_1\b W_1=-1/2\p_{\mu}\p_{\nu}h^{\mu\nu}+\p_{\mu}\p^{\mu}(1/2h+2\Phi_1).
\end{eqnarray}
The second order equations are more complicated. Again, we start from \rf{comp2}, namely, using the properties of 
the operator products we will rewrite it in the following way (from now on, we will omit zero index in the operator products):
\begin{eqnarray}\label{sa}
&&(L_0V_2)-V_2-1/2(V_1,V_1)^{(1,1)}+( \b W_1,V_1)^{(0,1)}+(W_1,V_1)_0^{(1,0)}-\nonumber\\
&&+\b L_{-1}W'_2+ L_{-1} \b W'_2=0,\\
&&\b W'_2(z)=-1/2((L_1 V_2)(z)-(L_1V_1+\b L_{-1} U_1,V_1)^{(1,1)}(z)-\nonumber\\
&&(\b L_1V_1+L_{-1} \b U_1,V_1)^{(2,0)}(z)+(U_1,V_1)^{(1,0)}(z)+\b L_{-1}U'_2(z)),\nonumber\\
&&W'_2(z)=-1/2((\b L_1 V_2)(z)-(\b L_1V_1+L_{-1} \b U_1,V_1)^{(1,1)}(z)-\nonumber\\
&&(L_1V_1+\b L_{-1} U_1,V_1)^{(0,2)}(z)+(\b U_1,V_1)^{(0,1)}(z)+ L_{-1}\b U'_2(z)),
\end{eqnarray}
where 
\begin{eqnarray}
U'_2=U_2-1/2(V_1,V_1)^{(2,2)}-1/2(U_1,V_1)^{(1,1)}+1/2(L_1V_1+\b L_{-1}U_1,V_1)^{(1,2)},\nonumber\\
\b U'_2=\b U_2-1/2(V_1,V_1)^{(2,2)}-1/2(\b U_1,V_1)^{(1,1)}+1/2(\b L_1V_1+L_{-1}\b U_1,V_1)^{(2,1)}.
\nonumber
\end{eqnarray}
 $W$-terms are:
\begin{eqnarray}\label{first}
&&W'_2=1/8\p^{\xi}(h_{\alpha\beta}\eta^{\beta\nu}h_{\nu\xi}+2s_{\xi\alpha})\p X^{\alpha}+\\ 
&&1/8(\p^{\beta}h_{\beta\xi}-2\p_{\xi}(U_1+\b U_1))
\eta^{\xi\rho}h_{\rho\alpha} \p X^{\alpha}-1/2\p_{\alpha}\b U'_2\p X^{\alpha}+O(\alpha'),\nonumber\\
&&\b W'_2=1/8\p^{\xi}(h_{\alpha\beta}\eta^{\beta\nu}h_{\nu\xi}+2s_{\xi\alpha})\bp X^{\alpha}+\\
&&1/8(\p^{\beta}h_{\beta\xi}-2\p_{\xi}(U_1+\b U_1))\eta^{\xi\rho}h_{\rho\alpha} \bp X^{\alpha}-
1/2\p_{\alpha} U'_2\bp X^{\alpha}+O(\alpha')\nonumber.
\end{eqnarray}
Here are the explicit formulas for other terms in sum \rf{sa}:
\begin{eqnarray}
&&(L_0-1)V_2=(L_0-1)(1/2\alpha'^{-1}s_{\mu\nu}\p X^{\mu}\bp 
X^{\nu})+\nonumber\\
&&(L_0-1)(1/4\alpha'^{-1}h_{\mu\beta}\eta^{\beta\alpha}h_{\nu\alpha}\p X^{\mu}\bp 
X^{\nu}),\\
&&(L_0-1)(1/2\alpha'^{-1}s_{\mu\nu}\p X^{\mu}\bp 
X^{\nu})=-1/4\Delta s_{\mu\nu}\p X^{\mu}\bp X^{\nu}, \\
&&(L_0-1)(1/4\alpha'^{-1}h_{\mu\beta}\eta^{\beta\alpha}h_{\nu\alpha}\p X^{\mu}\bp 
X^{\nu})=\\
&&-1/8\Delta h_{\mu\nu}\eta^{\nu\alpha}h_{\alpha\beta}\partial X^{\mu}
\bar{\partial}X^{\beta}-
1/8h_{\mu\nu}\eta^{\nu\alpha}\Delta h_{\alpha\beta}\partial X^{\mu}\bar{\partial}X^{\beta}\nonumber\\
&&-1/4\eta^{\xi\rho}\partial_{\xi}h_{\mu\nu}\eta^{\nu\alpha}\p_{\rho}h_{\alpha\beta}
\p X^{\mu}\bp X^{\beta}=-1/4\eta^{\xi\rho}\partial_{\xi}h_{\mu\nu}\eta^{\nu\alpha}\p_{\rho}h_{\alpha\beta}
\p X^{\mu}\bp X^{\beta}\nonumber\\
&&-1/8(\p_{\nu}\p^{\xi}h_{\xi\mu}\eta^{\nu\alpha}h_{\alpha\beta}+
\p_{\nu}\p^{\xi}h_{\xi\beta}\eta^{\nu\alpha}h_{\alpha\mu})
\partial X^{\mu}\bp X^{\beta}-\nonumber\\
&&1/8\p_{\nu}((\p^{\beta}h_{\beta\xi}-2\p_{\xi}(U_1+\b U_1))
\eta^{\xi\rho}h_{\rho\alpha} )\bp X^{\nu} \p X^{\alpha}-\nonumber\\
&&1/8\p_{\nu}((\p^{\beta}h_{\beta\xi}-2\p_{\xi}(U_1+\b U_1))
\eta^{\xi\rho}h_{\rho\alpha} )\p X^{\nu} \bp X^{\alpha}+\nonumber\\
&&1/8(\p^{\beta}h_{\beta\xi}-2\p_{\xi}(U_1+\b U_1))
\eta^{\xi\rho}\p_{\nu}h_{\rho\alpha}\bp X^{\nu} \p X^{\alpha}+\nonumber\\
&&1/8(\p^{\beta}h_{\beta\xi}-2\p_{\xi}(U_1+\b U_1))
\eta^{\xi\rho}\p_{\alpha}h_{\rho\nu}\bp X^{\nu} \p X^{\alpha},\nonumber
\end{eqnarray}
\begin{eqnarray}
&&(V_1,V_1)^{(1,1)}=\\
&&(4\alpha'^2)^{-1}(h_{\rho\sigma}\p X^{\rho}\bp X^{\sigma}, h_{\lambda\mu}\p X^{\lambda}\bp X^{\mu})^{(1,1)}=\nonumber\\
&&1/2\eta^{\xi\rho}\eta^{\sigma\alpha}h_{\rho\sigma}\p_{\xi}\p_{\alpha}h_{\mu\nu}
\p X^{\mu}\bp X^{\nu}-
1/2\eta^{\xi\rho}\eta^{\lambda\alpha}\p_{\lambda}h_{\xi\nu}\p_{\rho}h_{\alpha\sigma}\p X^{\nu}\bp X^{\sigma}-
\nonumber\\
&&1/4\p_{\sigma}\p_{\rho}h_{\mu\nu}h_{\xi\alpha}\eta^{\xi\nu}\eta^{\sigma\alpha}\p X^{\mu}\bp X^{\rho}-
1/4\p_{\sigma}\p_{\rho}h_{\mu\nu}h_{\xi\alpha}\eta^{\xi\nu}\eta^{\sigma\alpha}\p X^{\rho}\bp X^{\mu}+\nonumber\\
&&1/4\p_{\rho}h_{\mu\nu}\p_{\sigma}h_{\xi\alpha}\eta^{\xi\nu}\eta^{\sigma\mu}\p X^{\rho}\bp X^{\alpha}+
1/4\p_{\rho}h_{\mu\nu}\p_{\sigma}h_{\xi\alpha}\eta^{\xi\nu}\eta^{\sigma\mu}\p X^{\alpha}\bp X^{\rho}+\nonumber\\
&&1/4h_{\alpha\rho}\p_{\nu}\p_{\mu}h_{\xi\lambda}\eta^{\alpha\xi}\eta^{\rho\lambda}\p X^{\mu}\bp X^{\nu}+O(\alpha'),
\nonumber
\end{eqnarray}
\begin{eqnarray}
&&-1/2(\b L_1V_1+L_{-1}\b U_1,V_1)^{(1,0)}=\\
&&(8\alpha')^{-1}((\p^{\beta}h_{\beta\xi}-2\p_{\xi}\b U_1)\p X^{\xi}, h_{\mu\nu} \p X^{\mu}\bp X^{\nu})^{(1,1)}=\nonumber\\
&&-1/8\p_{\rho}(\p^{\beta}h_{\beta\xi}\eta^{\xi\alpha}h_{\alpha\beta})\p X^{\rho}\bp X^{\beta}+
1/8\p^{\beta}h_{\beta\xi}\eta^{\xi\alpha}\p_{\rho}h_{\alpha\beta}\p X^{\rho}\bp X^{\beta}+\nonumber\\
&&1/8\p_{\lambda}\p^{\beta}h_{\beta\rho}\eta^{\lambda\alpha}h_{\alpha\beta}\p X^{\rho}\bp X^{\beta}
-1/8(\p^{\beta}h_{\beta\xi}-2\p_{\xi}\b U_1)\eta^{\xi\lambda}\p_{\lambda}h_{\rho\beta}\p X^{\rho}\bp X^{\beta}+\nonumber\\
&&O(\alpha'),
\nonumber
\end{eqnarray}
\begin{eqnarray}\label{last}
&&-1/2(L_1V_1+\b L_{-1}U_1,V_1)^{(0,1)}=\\
&&(8\alpha')^{-1}((\p^{\beta}h_{\beta\xi}-2\p_{\xi}U_1)\bp X^{\xi}, h_{\mu\nu} \p X^{\mu}\bp X^{\nu})^{(1,1)}=
\nonumber\\
&&-1/8\p_{\rho}(\p^{\beta}h_{\beta\xi}\eta^{\xi\alpha}h_{\alpha\beta})\bp X^{\rho}\p X^{\beta}+
1/8\p^{\beta}h_{\beta\xi}\eta^{\xi\alpha}\p_{\rho}h_{\alpha\beta}\bp X^{\rho}\p X^{\beta}+\nonumber\\
&&1/8\p_{\lambda}\p^{\beta}h_{\beta\rho}\eta^{\lambda\alpha}h_{\alpha\beta}\bp X^{\rho}\p X^{\beta}
-1/8(\p^{\beta}h_{\beta\xi}-2\p_{\xi}U_1)\eta^{\xi\lambda}\p_{\lambda}h_{\rho\beta}\bp X^{\rho}\p X^{\beta}+\nonumber\\
&&O(\alpha').\nonumber
\end{eqnarray} 
Collecting formulae \rf{first}-\rf{last} in \rf{sa}, we arrive to the Einstein equations \rf{s}, putting 
\begin{eqnarray}
U_2+\b U_2=1/2s+2\Phi_2+1/4h_{\mu\nu}h^{\mu\nu}=U'_2+\b U'_2+1/8h_{\mu\nu}h^{\mu\nu}+O(\alpha').
\end{eqnarray}
Now let's obtain the last equation \rf{phi2} from \rf{dil2}. Let's write the expressions for different terms 
from equation \rf{dil2}:
\begin{eqnarray}
&&W_2=1/8\p^{\xi}(h_{\alpha\beta}\eta^{\beta\nu}h_{\nu\xi}+2s_{\xi\alpha})\p X^{\alpha}+\nonumber\\
&&1/16(\p^{\beta}h_{\beta\xi}-\p_{\xi}(2U_1+4\b U_1))\eta^{\xi\rho}h_{\rho\alpha}\p X^{\alpha}+\nonumber\\
&&1/32\p_{\rho}(h_{\mu\nu}h^{\mu\nu})-1/2\p_{\rho}\b U_2\p X^{\rho},
\end{eqnarray}
\begin{eqnarray}\label{d1}
&&{\alpha'}^{-1}(2(L_1W_2)-2(L_0U_2))=-1/4\p_{\xi}\p_{\alpha}(h^{\xi\mu}\eta_{\mu\nu}h^{\nu\alpha}+2s^{\xi\alpha})-\nonumber\\
&&1/16\p_{\alpha}\p^{\alpha}(h^{\mu\nu}h_{\mu\nu})+\p_{\alpha}\p^{\alpha}(U_2+\b U_2)-\nonumber\\
&&1/8(\p_{\alpha}\p^{\beta}h_{\beta\xi}-\p_{\alpha}\p_{\xi}(2U_1+4\b U_1))h^{\alpha\xi}-\nonumber\\
&&1/8(\p^{\beta}h_{\beta\xi}-\p_{\xi}(2U_1+4\b U_1))\p_{\alpha}h^{\nu\xi}=\nonumber\\
&&-1/2\p_{\xi}\p^{\alpha}h^{\xi\mu}h_{\mu\alpha}-1/4\p_{\xi}h^{\xi\mu}\p^{\alpha}h_{\mu\alpha}-\nonumber\\
&&1/4\p^{\alpha}h^{\xi\mu}\p_{\xi}h_{\mu\alpha}-1/2\p_{\xi}\p_{\alpha}s^{xi\alpha}-\nonumber\\
&&1/8h^{\alpha\xi}\p_{\alpha}\p^{\beta}h_{\beta\xi}+1/8h^{\alpha\xi}\p_{\alpha}\p_{\xi}(2U_1+4\b U_1)-\nonumber\\
&&1/8\p_{\alpha}h^{\alpha\xi}\p^{\beta}h_{\beta\xi}+1/8\p_{\xi}(2U_1+4\b U_1)\p_{\alpha}h^{\alpha\xi}+\nonumber\\
&&1/2\p_{\mu}\p^{\mu}(1/2s+2\Phi_2)+3/4\p^{\rho}\p_{\nu}h^{\nu\alpha}h_{\rho\alpha}+\nonumber\\
&&3/8\p^{\rho}h_{\mu\nu}\p_{\rho}h^{\mu\nu}-3/4h^{\mu\nu}\p_{\mu\nu}(U_1+\b U_1),
\end{eqnarray}
\begin{eqnarray}\label{d2}
&&{\alpha'}^{-1}2(U_1,W_1)^{(1,0)}=1/2(\p^{\beta}h_{\beta\xi}-2\p_{\xi}\b U_1)\p^{\xi}U_1+O(\alpha')\nonumber\\
&&{\alpha'}^{-1}2(W_1,W_1)^{(2,0)}=-1/8(\p^{\beta}h_{\beta\xi}-2\p_{\xi}\b U_1)(\p_{\beta}h^{\beta\xi}-2\p^{\xi}\b U_1)+O(\alpha')\nonumber\\
&&\alpha'^{-1}(U_1,\b W_1)^{(0,1)}=1/4(\p^{\beta}h_{\beta\xi}-2\p_{\xi}U_1)\p^{\xi}U_1+O(\alpha')\nonumber\\
&&-\alpha'^{-1}(V_1,W_1)^{(2,1)}=-1/8h^{\xi\alpha}\p_{\alpha}(\p^{\beta}h_{\beta\xi}-2\p_{\xi}\b U_1)+O(\alpha')\nonumber\\
&&\alpha'^{-1}(V_1,U_1)^{(1,1)}=1/2h^{\alpha\beta}\p_{\alpha}\p_{\beta}U_1.
\end{eqnarray}
Summing \rf{d1} and \rf{d2}, we arrive to  equation \rf{dil2}. 
It is easy to see that for our choice of $V$ and $U, \b U$-terms, \rf{dil22} leads to the same equation. 
This ends the proof of  the proposition.$\blacksquare$

\end{document}